\documentclass[mathleft,fleqn,%
]{an}
\usepackage{graphicx}
\usepackage[varg]{txfonts}
\pdfoutput=1
\begin{document}

\Pagespan{1}{}
\Yearpublication{2014}%
\Yearsubmission{2014}%
\Month{0}%
\Volume{999}%
\Issue{0}%
\DOI{asna.201400000}%

\title{Gaia}

\author{C. Cacciari\thanks{Corresponding author:
        {carla.cacciari@oabo.inaf.it}}, E. Pancino \and M. Bellazzini 
}
\titlerunning{Gaia}
\authorrunning{C. Cacciari et al. }
\institute{
INAF, Osservatorio Astronomico di Bologna, Italy}

\received{XXXX}
\accepted{XXXX}
\publonline{XXXX}


\abstract{A review of the Gaia mission and its science performance after one year of operations 
will be presented, and the contribution to reconstructing the history of the Milky Way will 
be outlined.  }

\maketitle

\section{Introduction}\label{s:intro}

Gaia is an ESA cornerstone mission building on the Hipparcos heritage (Perryman et al. 1997; 
van Leeuwen 2007).  
As stated in  ``Gaia - Composition, Formation and Evolution of the Galaxy: Concept
and Technology Study Report\footnote{Gaia Science Advisory Group, ESA-SCI 2000-4, (2000)}:
{\it The primary objective of Gaia is the Galaxy: to observe the physical
characteristics, kinematics and distribution of stars over a large fraction
of its volume, with the goal of achieving a full understanding of the MW
dynamics and structure, and consequently its formation and history}. 
  
The satellite and payload were commissioned to the industry (Airbus Defence \& Space), 
the management, launch and operations are performed by the European Space Agency (ESA), and the 
data processing is entrusted to a Pan-European cooperation represented by the Data Processing 
and Analysis Consortium (DPAC) which includes about 450 scientists from more than 20 countries. 

Gaia was launched on 19 December 2013 from Kourou (French Guiana) with a Soyuz-Fregat rocket, 
and was put on a Lissajous orbit around L2, 1.5 million km away from the Sun, for an expected 
5yr lifetime of operations. It will perform an all sky survey complete to G$\sim$20.7, 
corresponding to V $\sim$ 20-22 mag for blue and red objects respectively, 
obtaining astrometric data with micro arcsec ($\mu$as) accuracy for more than 10$^9$ objects, 
as well as optical spectrophotometry for most of them, and medium resolution spectroscopy at 
the Ca II triplet (847-870 nm) for several million objects brighter than 16th mag. 

Gaia data have no proprietary rights, they will be released at planned intervals after proper 
validation.  The final Catalogue is foreseen for 2022, but there will be intermediate data 
releases. The first one is scheduled for summer 2016, and will contain positions and white-light 
photometry for single stars from $\sim$90\% of the sky, data from the Ecliptic Pole regions that 
were scanned more often during commissioning, and proper motions for the about hundred thousand 
Hipparcos stars. It is presently under evaluation whether to deliver positions, parallaxes and 
proper motions for a fraction of the 2.5 million Tycho-2 stars (H\o g et al. 2000). 
These astrometric data will be obtained by combining the positions from the Tycho-2 Catalogue 
with the first year of Gaia data for a joint Tycho-Gaia astrometric solution (TGAS, Michalik et 
al. 2015; see also the Gaia webpage\footnote{http://www.cosmos.esa.int/web/gaia/news\_20150807.}). 
Thanks to the $\sim$24 yr time baseline, these data will have sub-mas accuracy. 
The second data release is currently planned for early 2017, to deliver radial velocities for 
bright stars, two-band photometry, and full astrometric data where available. 
Two more data releases are planned around  2018-19. Each release 
updates the previous ones and contains significant new additions. 
Science alerts (e.g. SNe) data are released immediately after validation. 

Gaia represents a huge improvement with respect to Hipparcos: the detection limit is
extended by $\ge$8 magnitudes, the number of sources is 10$^4$ times larger and includes objects
unobservable by Hipparcos such as galaxies and quasars, and the astrometric accuracy is 
a factor $\sim$ 100 better (see Table \ref{t:h2g}   and the Gaia 
webpage\footnote{http://www.cosmos.esa.int/web/gaia.} for more details.). 

\begin{table}
\caption{From Hipparcos to Gaia.}
\label{t:h2g}  
\centering     
\begin{tabular}{lll}\hline
    & Hipparcos &  Gaia  \\
\hline
 Magnitude limit   & V$_{lim}$ = 12 & V$_{lim}$ $\sim$ 20-22  \\
 N. of objects     & 1.2$\times$10$^5$  & $\ge$10$^9$  \\
 Quasars           & none & $\sim$5$\times$10$^5$ \\
 Galaxies          & none & $\sim$ 10$^6$-10$^7$ \\
 Astrom. accuracy  & $\sim$1 mas & 5-14$\mu$as at V$\le$12 \\
                   &             & 20-30$\mu$as at V=15, \\
		   &             & 400-600$\mu$as at V=20 \\
 Broad-band phot.  & 2 (B,V) & 3 (to V$_{lim}$) \\ 
                   &         & + 1 (to V=16) \\
 Spectrophtometry  & none & 2 bands (B/R) to V$_{lim}$ \\
 Spectroscopy      & none & 1-15 km/s to V=16 \\
 Obs. programme    & pre-selected  & all-sky complete  \\
                   & targets      & and unbiased \\
\hline 
\end{tabular}
\end{table}

\section{Technicalities in a nutshell}\label{s:nut}

\subsection{Satellite}\label{ss:satellite}
The satellite spins around its axis, which is oriented 45-deg away from the Sun,
with a period of 6 hr, and the spin axis has a precession motion around the solar
direction with a period of 63 days. The combination of these two motions results in
a quasi-regular time-sampling that allows to scan the entire sky on average 70 times 
over the 5 yr mission lifetime, with as much as $\sim$200 transits in a $\pm$ 10-deg 
strip around ecliptic latitudes  $\pm$45deg, and as little as  $\sim$50 transits 
in other areas of the sky. 

\subsection{Payload}\label{ss:payload}
Global (wide field) astrometry requires repeated observation of the same area of the sky from 
two lines of sight separated by a suitable and very stable angle (Basic Angle, BA), producing 
two fields of view which get combined on the focal plane. 
Gaia can then make relative measurements among the stars simultaneously visible in the combined 
field of view. 
To this purpose, the payload is a toroidal structure holding two primary 1.45m$\times$0.50m 
rectangular mirrors (field of view FoV = 1.7-deg$\times$0.6-deg) whose lines of sight are separated
by a BA angle of 106.5-deg. The BA needs to be stable to an extremely high 
degree to ensure Gaia's expected astrometric accuracy\footnote{For the measurement of stellar 
parallaxes down to 10 $\mu$as accuracy, the requirement for the BA stability would be 1 part 
in 3$\times$10$^{10}$ (van Leeuwen 2007).}.  Therefore
a BA monitoring system is hosted on the payload, along with all the optical components 
which allow to combine the FoVs of the two mirrors on the focal plane. 
The toroidal structure hosts also the prisms for the low resolution 
spectrophotometry, the radial velocity spectrometer (RVS) and the array of CCD detectors 
on the focal plane.  

\subsection{Focal Plane }\label{ss:focplane} 
The focal plane contains several arrays of 4.5K$\times$2.0K CCDs: 
{\bf i)} the sky mapper (SM), 2$\times$7 CCDs for detection and confirmation of 
source transit; 
{\bf ii)} the astrometric field (AF), 9$\times$7 CCDs corresponding to 
40$\times$40 arcmin on the sky, for astrometric measurements and white light 
(G-band, 330-1050 nm) photometry; 
{\bf iii)} the blue (BP) and red (RP) photometers, 2$\times$7 CCDs for low 
resolution (R$<$100) slitless prism spectro-photometry in the ranges 330-680 nm 
and 640-1050 nm, respectively. From the spectra the integrated G$_{BP}$ and 
G$_{RP}$ magnitudes (and colour) are derived.
{\bf iv)} the radial velocity spectrometer (RVS), 3$\times$4 CCDs for slitless 
spectroscopy at the Ca II triplet (845-872 nm) with R$\sim$11,500.

Measurements are made in time-delayed-integration mode, reading the CCDs
at the same speed as the sources trail across the focal plane, i.e. 
60 arcsec/sec, corresponding to a crossing/reading time of 4.4 sec per CCD.

\section{Science Performance}\label{s:sperf} 
The commissioning phase was completed on 18 July 2014. The overall performance 
was assessed to be very good, and the detection efficiency was improved so as to extend 
the bright end to G$\sim$0 mag through detection algorithm improvements and employment of a
special observing mode. The faint limit was also extended down to G=20.7 mag. 
However, a few problems were identified: stray light (mostly due to scattering from filaments 
on sun-shield edges), contamination (thin ice layer on optics), some larger than expected 
variations of the BA. 
These effects can be mitigated (e.g. by periodic de-contamination campaigns) or modelled, and 
actions have been taken to this purpose. A performance reassessment was made after careful 
analysis of the commissionig data, and the updated characteristics as of March 2015 are summarized 
in the following  
sub-sections\footnote{See http://www.cosmos.esa.int/web/gaia/science-performance for more details.}.  

\subsection{Astrometry}\label{ss:astroperf}

Astrometric errors are dominated by photon statistics. 
The end-of-mission astrometric performance estimated by de Bruijne (2015) is  summarized in 
Table \ref{t:aperf}.  The range of values for the stars brighter than 
12th mag is mostly due to bright-star observing conditions (e.g. TDI gates, onboard 
mag-estimation errors, etc.). 
We remind that the standard errors on position and proper motion are about 0.74 and 0.53 of 
those on parallax, respectively.

\begin{table}[h]
\caption{Predicted sky-averaged end-of-mission standard errors on the parallax as a function 
of Johnson V magnitudes, in units of $\mu$as, for unreddened  B1V, G2V and M6V spectral types. }
\label{t:aperf}
\centering
\begin{tabular}{c|c|c|c}
\hline
V[mag]  &  B1V &  G2V &  M6V   \\
\hline
$\le$12   &  5-14 & 5-14 & 10  \\
15        &  26   & 24   & 9   \\ 
20        &  600  & 540  & 130 \\
\hline
\end{tabular}
\end{table} 

Gaia's first observations of gravitational lensed images of two QSOs, Q2237+030 (Einstein Cross) 
and HE0435-1223\footnote{See http://www.cosmos.esa.int/web/gaia/iow$\_$20150409, Gaia image of 
the week 9 April 2015.}, provide an example of the excellent astrometric quality obtained already 
with preliminary data and reduction process: the two QSOs are located at z=1.7 and the lensed images 
have V $\sim$ 17-19 mag. Even with a very small number of observations the accuracy of the 
absolute position for each of the Einstein Cross images is $\sim$ 50 mas, and slightly better 
for HE0435-1223 for which more observations are available. 
These positions were obtained from the Initial Data Treatment in a routine mode, with a 
very preliminary attitude determination. By end-2015 nine new observations of Q2237+030 
and sixteen of HE0435-1223 will be obtained, and a much better accuracy will be achieved thanks also 
to the use of the improved attitude determination.

\subsection{Photometry}\label{ss:photoperf}
Gaia photometry includes the integrated white-light (G-band)
from the AF, and the BP/RP prism spectra from which the G$_{BP}$ and G$_{RP}$ integrated
magnitudes are derived. The expected end-of-mission errors are shown in Table \ref{t:phperf} 
(for the relations between the G magnitudes and other photometric systems see 
Jordi et al. 2010). 

For about 28 days in July 2014 Gaia scanned continuously a $\sim$1-deg area at the North 
and South Ecliptic Poles, collecting a large number of  measurements for several hundreds 
RR Lyraes and Cepheids in the Large Magellanic Cloud, a good fraction of which were new 
discoveries. The light curves of the RR Lyraes in the range 18.5-19.5 mag, and of the 
Cepheids in the range 18-19 mag, are presented in the 5 March and 28 May 2015 
``Gaia image of the week'' section\footnote{See http://www.cosmos.esa.int/web/gaia/iow$\_$20150305 
and http://www.cosmos.esa.int/web/gaia/iow$\_$20150528, respectively.}, and show the excellent 
photometric quality obtained even at the faint end of the dynamic range. 

Astrophysical parameters can be derived using BP/RP spectral energy distributions, 
sometimes in combination with astrometric and spectroscopic data. Preliminary simulations 
by Bailer-Jones et al. (2013) on F, G, K, M dwarfs and giants for a wide range of metallicites
and interstellar extinctions indicate the following internal accuracy (rms residuals) of 
astrophysical parameter estimates, to be reassessed on real data: at magnitude G=15 (i.e. 
V$\sim$15-17 depending on the source colour) the temperature can be derived to $\sim$100 K, 
the extinction A$_V$ to $\sim$0.1 mag, the gravity to $\sim$0.25 dex and the metallicity to 
$\sim$0.2 dex. The accuracy is a strong function of the parameters themselves, and may vary 
by a factor of more than two up or down over this parameter range.

\begin{table}[h]
\caption{End-of-mission expected standard photometric errors as a function of Gaia G magnitude, 
in the G/G$_{BP}$/G$_{RP}$ bands, in units of milli-magnitude, for  B1V, G2V and M6V spectral 
types. }
\label{t:phperf}
\centering
\begin{tabular}{c|c|c|c}
\hline
G[mag]  &  B1V &  G2V &  M6V   \\
\hline
15     & 1/4/4  & 1/4/4  & 1/7/4 \\ 
18     &  2/8/19  & 2/13/11 &  2/89/6   \\ 
20     & 6/51/110 & 6/80/59 &  6/490/24 \\
\hline
\end{tabular}
\end{table}

\subsection{Spectroscopy}\label{ss:spectroperf}

The RVS provides the third component of the space velocity for sources down to about 16th 
magnitude. Radial velocities are the main product of the RVS, with typical end-of-mission 
errors of 1 (15) km/sec at V$\sim$7.5 (11.3) for a B1V star, at V$\sim$12.3 (15.2) for a 
G2V star, and  at V$\sim$12.8 (15.7) for a K1III metal-poor star. 
For sources brighter than  $\sim$12 mag the RVS spectra will provide information also on
rotation and chemistry, and combined with the prism BP/RP spectra will allow us to
obtain more detailed and accurate astrophysical parameters.

An example of the first RVS data obtained by Gaia is shown in the ``Gaia image of the week'' 
on 26 January 2015\footnote{See http://www.cosmos.esa.int/web/gaia/iow$\_$20150126.}, where 
high signal-to-noise spectra of three bright hot stars show very well defined features, 
in particular diffuse interstellar bands which, in combination with the Gaia parallaxes, 
will allow the construction of an unprecedented 3D map of the interstellar medium.

\section{Science products}\label{s:science} 

Gaia will provide a complete census of all Galactic stellar populations, as 
well as  extragalactic sources, down to 20th magnitude. 
This wealth of data will have a major impact in several areas of astronomy 
and astrophysics, e.g.: \\
$\bullet$ {\it Galactic studies}: spatial and dynamical structure, formation and chemical history 
of all MW components. Based on the Besan\c con Galaxy model (Robin et al. 2012) 
Gaia will observe about 9$\times$10$^8$ stars in the disk(s), 2$\times$10$^7$ in the spheroid 
and 1.7$\times$10$^8$ in the bulge. \\
Proper motions will allow typical tracers such as K-M giants to map the kinematic 
characteristics of the bulge to about 2.5 kms$^{-1}$, of the thin disk to about 1-2 kms$^{-1}$ 
as far as 10 kpc, of the thick disk to $\le$5 kms$^{-1}$  as far as 15 kpc, and of 
the halo to better than 1 (5) kms$^{-1}$ as far as 10 (30) kpc. 
Also horizontal branch (A5) stars will be able to characterize these MW components with 
an accuracy 5 up to 100 times better than their typical velocity dispersion. 
Blue/red supergiants and Cepheids will characterize the spiral arms to 0.2-2.0 kms$^{-1}$ 
as far as 10 kpc.  All the open clusters will be observed with a large enough number of 
stars to derive very accurate mean distances, and for the stars brighter than 16th mag 
the astrophysical parameters will be estimated. \\
$\bullet$ {\it  Cosmic Distance scale}:  Gaia is expected to observe a large number of standard 
candles in the Galaxy, e.g. RR Lyrae stars (as many as $\sim$ 4$\times$10$^4$ in the bulge and 
7$\times$10$^4$ in the halo) with $\le$1 (5)\% accuracy within 2 (4) kpc, and Cepheids 
(2-9$\times$10$^3$  -  Eyer \& Cuypers, 2000; Windmark et al. 2011) with $\le$1\% accuracy for 
about 1/3 of the total sample and  $\le$5\% accuracy for the entire sample. This will provide 
the first trigonometric calibration of local candles for a reliable and accurate 
definition of the cosmological distance scale; \\
$\bullet$ {\it Stellar Physics}: accurate distances will be particularly useful as a check of the 
asteroseismic scaling laws, and  with the additional help from photometry and astrophysical 
parameters will provide a powerful benchmark for a better understanding of stellar structure and 
evolution modelling; \\ 
$\bullet$ Several other areas will greatly benefit from Gaia's data: Solar System Objects 
(a few 10$^5$), exoplanets  (a few 10$^4$), fundamental physics 
(general relativity experiments), gravitational lensing events (a few 10$^2$), 
brightest stellar populations in nearby (LG) galaxies, supernovae and burst sources 
(a few 10$^4$), galaxies (10$^6$ - 10$^7$) and QSOs (a few 10$^5$), definition 
of the International Celestial Reference Frame (ICRF). 

\section{Stellar systems as observed by Gaia} 

As an example, it is interesting to see how Gaia is expected to improve our understanding 
of stellar systems such as globular clusters (GC) in the Milky Way, and dwarf spheroidal 
(dsph) galaxies in the Local Group. These two cases have been simulated using the most recent
information on Gaia performance and are presented in the following.

\subsection{Globular clusters in the MW}\label{ss:gc}

A full 3-D synthetic GC at dynamical equilibrium, with assumed absolute 
integrated magnitude of M$_V$=-7.6, was costructed by Pancino  et al. (2013). 
Magnitudes were assigned to all stars from appropriate 12-gyr old and metal-poor 
stellar models, and a typical velocity dispersion of $\sim$ 10 kms$^{-1}$ was applied. 
The cluster was placed at the distances of 5 and 10 kpc in different positions on the sky, and 
reddening and contamination by field stars were added according to the Besan\c con Galactic
model.  A systemic proper motion of $-$5000 $\mu$as/yr was assumed in both RA and DEC, 
corresponding to $-$118.5 and $-$237 kms$^{-1}$ at 5 and 10 kpc distance, respectively.

\begin{figure}
\includegraphics[scale=0.28,angle=-90]{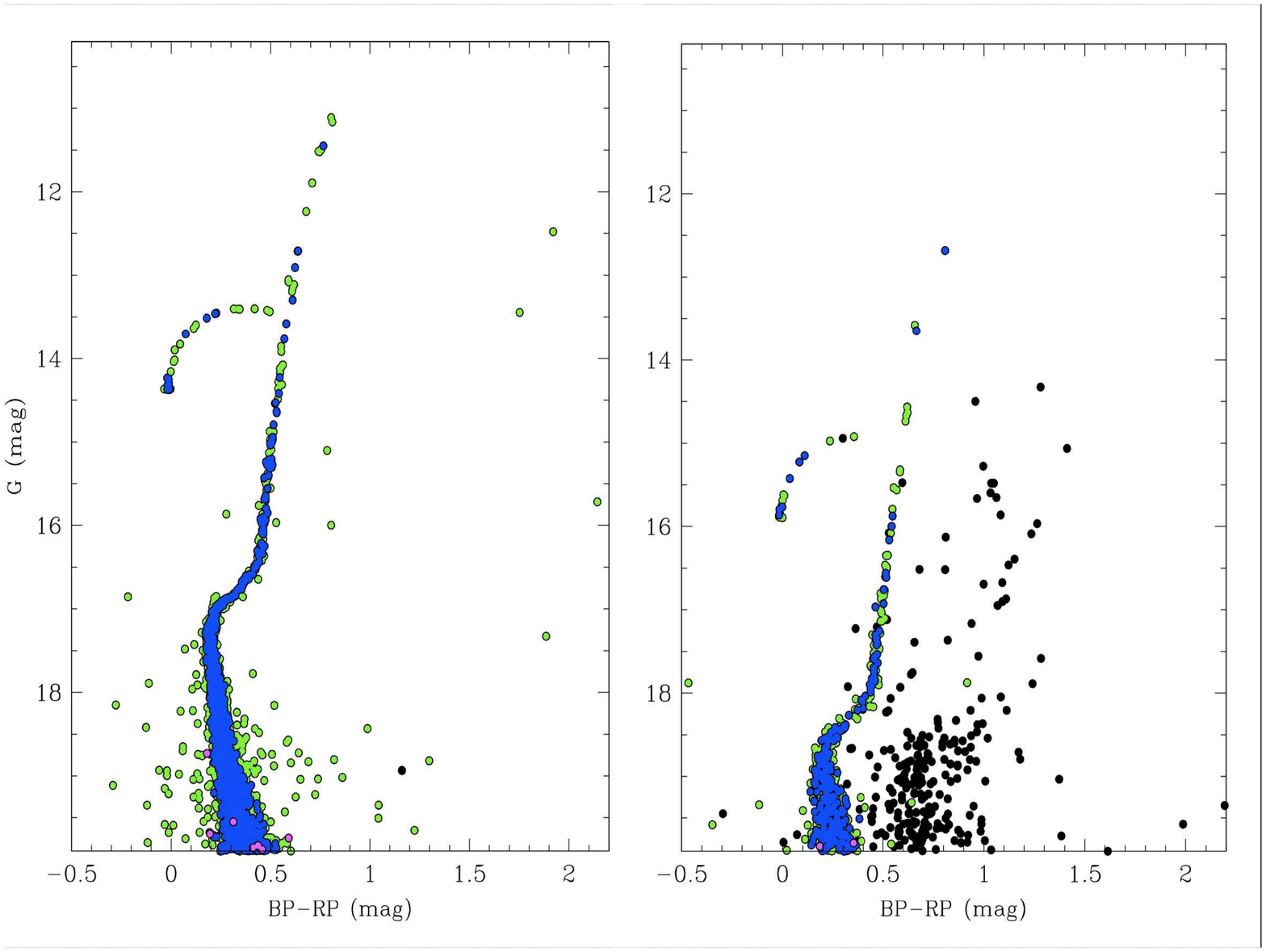}
\caption{Colour-magnitude diagram in Gaia photometry (BP-RP mag on X-axis, 
G$\le$20 mag on Y-axis) of a simulated globular cluster after field decontamination and deblending, 
at 5 kpc (left) and 10 kpc (right). See text for details. }
\label{f:gcc}    
\end{figure}

Figure 2 in Pancino et al. (2013) shows how the cluster colour-magnitude diagram  
looks like at 10 kpc (full sample). For completeness we show here in Figure \ref{f:gcc} the 
clean colour-magnitude diagrams at 5 and 10 kpc, after statistical field decontamination and 
deblending simulating Gaia observations and data processing, which retain only the well 
measured member stars brighter than 20th mag (16838 stars at 5 kpc and 3513 stars 
at 10 kpc). 

\begin{figure}
\includegraphics[scale=0.29,angle=90]{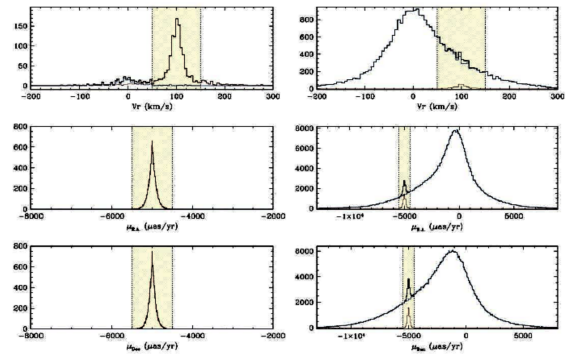}
\caption{Proper motion simulations of the globular clusters shown in 
Fig. \ref{f:gcc}:  at 5 kpc (left), at 10 kpc (right). }
\label{f:ep-pm}    
\end{figure}

Astrometric measures were derived for these stars, and end-of-mission errors from de Bruijne 
et al. (2015) were associated to these measures.  
The results of this simulation are presented in Table \ref{t:gc} and shown in 
Figure \ref{f:ep-pm}, where one can see that even at 10 kpc the cluster membership can be 
clearly disentangled and identified over a heavy background, and astrometric measures can be 
obtained with high accuracy and precision.    

%

%

\begin{table}[h]
\caption{Gaia simulated astrometric measures of the globular clusters shown in 
Fig. \ref{f:gcc}.}
\label{t:gc}       
\centering
\begin{tabular}{ll}\hline
   D = 5 kpc (200$\mu$as) &  D = 10 kpc (100$\mu$as) \\   
\hline
                   &                                  \\
 N. stars=16838   & N. stars=3513 \\
 pm(RA)=$-$4998.7$\pm$0.8 $\mu$as/yr    & pm(RA)=$-$4993$\pm$3 $\mu$as/yr \\
 pm(DEC)=$-$5000.2$\pm$0.7 $\mu$as/yr   & pm(DEC)=$-$4994$\pm$3 $\mu$as/yr \\
 $\pi$ = 199.7$\pm$0.7 $\mu$as      & $\pi$ = 101.2$\pm$1.4 $\mu$as  \\
 D = 5.007$\pm$0.007 kpc            & D = 9.997$\pm$0.017 kpc \\
\hline 
\end{tabular}
\end{table}

About 55\% of the Galactic globular clusters, i.e. $\sim$90 objects, are located within 
10 kpc from the Sun (Harris 1996, 2010 edition), and this simulation shows how important 
Gaia contribution will be in tracing and understanding the formation and evolution history 
of this Galactic component.  

\subsection{The Draco dwarf spheroidal galaxy}\label{ss:draco}

\begin{figure} 
\includegraphics[scale=0.35]{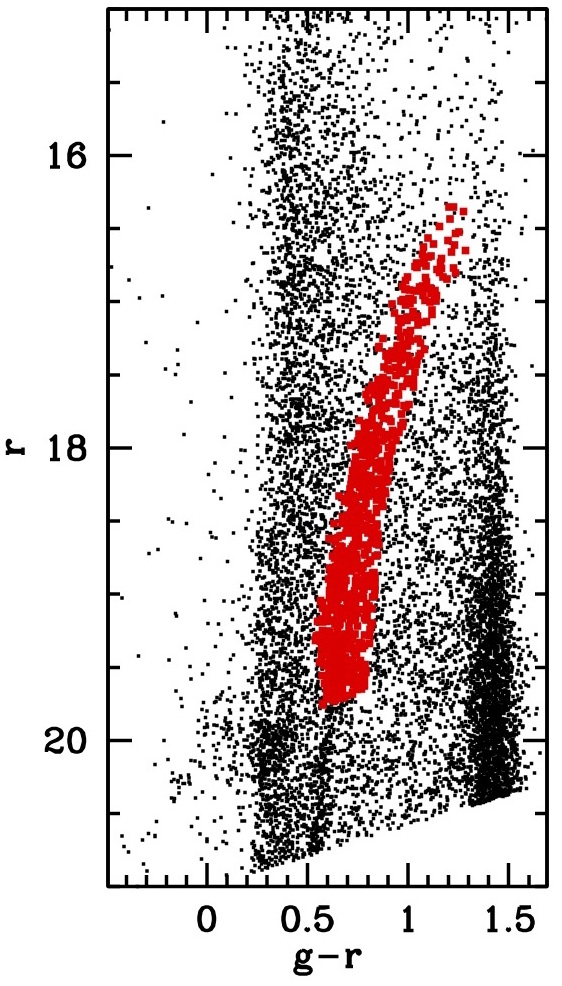}   
\caption{Colour-magnitude diagram of the Draco dsph galaxy: red dots indicate the 750 RGB stars 
brighter than 20th mag selected as bona fide members observable by Gaia. }
\label{f:cm-draco}    
\end{figure}

Using the CFHT photometric data by Segall et al. (2007) of a 2-deg$\times$2-deg area 
centered on the Draco dwarf galaxy (M$_V$=-9.0 mag, distance 93 kpc, central surface 
brightness of 25.0 mag/arcsec$^2$), we selected 750 red giant branch (RGB) 
stars with V$\le$20 mag as bona fide members after statistical field decontamination 
(see Fig. \ref{f:cm-draco}).  

%

\begin{figure}
\includegraphics[scale=0.35]{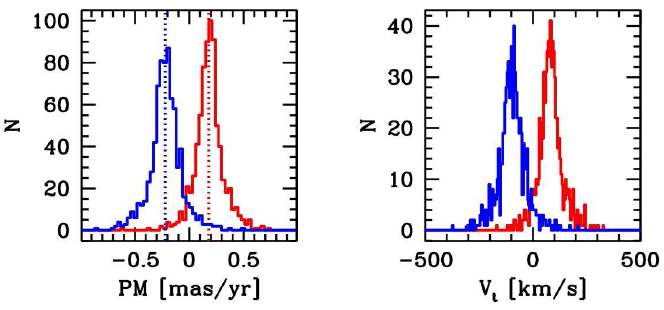}
\caption{Simulation  of the Draco proper motions, in units of mas/yr 
(left) and km/s (right): pm(RA) is plotted in red, pm(DEC) in blue.}
\label{f:drapm}    
\end{figure}

With a similar procedure as the one applied to GCs (Sect. \ref{ss:gc}), 
to each of these 750 stars we associated: i) a realistic systemic tangential motion, borrowed 
from the study of Draco proper motions using HST-ACS images by Pryor et al. (2015); 
ii) a realistic isotropic velocity dispersion of 9.1 kms$^{-1}$ (McConnachie, 2012); 
and iii) Gaia's observational errors from de Bruijne et al. (2015). \\ 
With these input data and assumptions,  and using a simple gaussian model, we obtained 
a Maximum Likelihood estimate of the systemic proper motion and velocity dispersion  
from the simulated Gaia measurements. By varying the integrated magnitude of the system from
M$_V$=-9.0 mag to -6.5 and -4.3 the number of member RGB stars decreases to 75 and 10,
respectively, and the simulation was performed also on these cases. 
The resulting proper motions are presented in Table \ref{t:dsph} columns 2-3 and shown 
in Fig. \ref{f:drapm}.  
The velocity dispersion $\sigma$-vel is estimated as 21.0$\pm$5.4$\mu$as/yr 
in the 750 star case, corresponding to 9.2$\pm$2.4 kms$^{-1}$, which recovers very well the 
input value of 9.1 kms$^{-1}$. The test cases with 75 and 10 stars have too few stars to 
produce reliable estimates of the  velocity dispersion. On the other hand, 
the systemic motion is reliably measured in all the cases. This 
can be fully appreciated by comparing the results of these simulations with Pryor et al.'s (2015) 
results, i.e.   pm(RA)=177$\pm$63 $\mu$as/yr and pm(DEC)=-221$\pm$63 $\mu$as/yr, which   
show that the Gaia end-of-mission astrometry is expected to perform about 
20 times better than the state of the art HST astrometry using 750 stars, and still 
two times better  using as few as 10 stars.
%
%
%
In any case, the synergy with the HST and many of the present and future surveys, especially 
the high-resolution spectroscopic ones, will be very important: none of them will provide 
all-sky coverage, but all will contribute to complement and extend the scientific potential 
of the Gaia data.

\begin{table}
\caption{Gaia simulated observations of the Draco dwarf spehroidal galaxy. 
Proper motions and velocity dispersion are in units of $\mu$as/yr.}
\label{t:dsph}       
\centering
\begin{tabular}{llll}\hline
   N. stars/M$_V$  &  pm(RA)          & pm(DEC)   &  $\sigma$-vel     \\  
\hline
  750/-9.0  & 175$\pm$3.3      & -217$\pm$3.1  & 21.0$\pm$5.4 \\ 
   75/-6.5  & 172$\pm$11.9     & -217$\pm$9.3  & 42.0$\pm$15.6 \\
   10/-4.3  & 180$\pm$30       & -185$\pm$30   & ---  \\	    
\hline 
\end{tabular}
\end{table}

\acknowledgements

This work is supported by the Istituto Nazionale di Astrofisica (INAF), and by the 
Agenzia Spaziale Italiana (ASI) through the grant 2014-025-R.0. 
CC gratefully acknowledges the support from the Wilhelm und Else Heraeus Foundation.


\begin{thebibliography}{} 
\bibitem{} Bailer-Jones, C.A.L., Andrae, R., Arcay, B., et al. 2013, \aap, 559, 74 
\bibitem{} de Bruijne, J.H.J., Ryg, K.L.J., \& Antoja, T. 2015, arXiv:150200791  
\bibitem{} Eyer, L. \&  Cuypers, J. 2000, {\it Astron. Soc. Pac. Conf. Ser.}, 203, 71 
\bibitem{} Harris, W.E. 1996, \aj, 112, 1487  
\bibitem{} H\o g, E., Fabricius, C., Makarov, V.V., et al. 2000, \aap, 355, L27 
\bibitem{} Jordi, C., Gebran, M., Carrasco, J.M., et al. 2010, \aap, 523, 48 
\bibitem{} McConnachie, A.W. 2012, \aj, 144, 4
\bibitem{} Michalik, D., Lindegren, L., \& Hobbs, D. 2015, \aap, 574, 115 
\bibitem{} Pancino, E., Bellazzini, M., \& Marinoni, S. 2013, Mem. S.A.It., 84, 83
\bibitem{} Perryman, M.A.C., Lindegren, L., Kovalevsky, J.,  et al. 1997, \aap, 323, L49 
\bibitem{} Pryor, C., Piatek, S. \& Olszewski, E.W. 2015, \aj, 149, 42 
\bibitem{} Robin, A.C., Luri, X., Reyl\'e, C., Isasi, Y., et al. 2012, arXiv:1202.0132 
\bibitem{} Segall, M., Ibata, R., Irwin, M.J., et al. 2007,  \mnras, 375, 831
\bibitem{} van Leeuwen, F. 2007, \apss ~Lib., Dordrecht:Springer, Vol. 350 
\bibitem{} Windmark, F., Lindegren, L., \& Hobbs, D. 2011, \aap, 530, 76 
\end{thebibliography}
\end{document}